\documentclass[12pt,a4paper,reqno]{amsart}
\usepackage{amssymb, amscd, amsthm, amsbsy, euscript}

\newtheorem{theor}{Theorem}
\theoremstyle{definition}
\newtheorem*{theorNo}{Theorem}

\newtheorem{state}[theor]{Proposition}

\newtheorem{define}{Definition}

\newtheorem{example}{Example}

\theoremstyle{remark}
\newtheorem{rem}{Remark}

\DeclareFontFamily{OML}{cyr}{}
\DeclareFontShape{OML}{cyr}{m}{n}{
   <5> <6> <7> <8> <9> gen * wncyr
   <10> <10.95> <12> <14.4> <17.28> <20.74> <24.88> wncyr10
  }{}
\DeclareSymbolFont{rusletters}{OML}{cyr}{m}{n}
\DeclareSymbolFontAlphabet{\rusmath}{rusletters}
\DeclareMathSymbol\re{\rusmath}{rusletters}{"03}
\newcommand{\cEv}{{\EuScript{E}}}   

\newcommand{\BBC}{\mathbb{C}}

\newcommand{\cE}{\mathcal{E}}

\newcommand{\cEToda}{{\mathcal{E}}_{\text{\textup{Toda}}}}
\newcommand{\cELiou}{{\cE}_{\text{\textup{Liou}}}}
\newcommand{\cH}{\mathcal{H}}

\newcommand{\cL}{\mathcal{L}}

\newcommand{\bu}{\boldsymbol{u}}
\newcommand{\gm}{\mathfrak{m}}

\newcommand{\gothg}{\mathfrak{g}}

\newcommand{\vph}{\varphi}
\newcommand{\dd}{\partial}
\newcommand{\Id}{{\mathrm d}}
\newcommand{\IL}{{\mathrm L}}
\newcommand{\cEL}{{\mathcal{E}}_{\IL}}

\DeclareMathOperator{\sym}{sym}

\DeclareMathOperator{\img}{im}

\newcommand{\ib}[3]{ \{\!\{ {#1},{#2} \}\!\}_{{#3}} }

\newcommand{\by}[1]{\textit{{#1}}}
\newcommand{\jour}[1]{\textit{{#1}}}
\newcommand{\vol}[1]{\textbf{{#1}}}
\newcommand{\book}[1]{\textrm{{#1}}}

\title[KdV\/-\/type systems for rank two simple Lie algebras: an illustration]%
{A geometric derivation of KdV\/-\/type hierarchies from root systems}

\author[A.~V.~Kiselev]{Arthemy V.~Kiselev}
\thanks{\textit{Address}:
Mathematical Institute, 
University of Utrecht, Budapestlaan~6, 3584~CD Utrecht, The
Netherlands.
\quad\textit{E-mails}: 
[\texttt{A.V.Kiselev}, \texttt{J.W.vandeLeur}]\texttt{\symbol{"40}uu.nl}}

\author[J.~W.~van de Leur]{Johan W. van de Leur}

\date{January 30, 2009}

\dedicatory{4th International workshop ``Group analysis of differential
equations and integrable systems'' \textup{(}Protaras, Cyprus, October 26--29, 2008\textup{)}}


\begin{document}
\begin{abstract}
For the root system of each complex semi\/-\/simple Lie algebra of rank two, and for the associated 2D~Toda chain $\cE=\bigl\{\bu_{xy}=\exp(K\bu)\bigr\}$, we calculate the two first integrals of the characteristic equation $D_y(w)\doteq0$ on~$\cE$. Using the integrals, we reconstruct and make coordinate\/-\/independent the $(2\times2)$-\/matrix operators~$\square$ in total derivatives that factor 
symmetries of the chains. Writing other factorizations that involve the operators~$\square$, we obtain pairs of compatible Hamiltonian operators
that produce KdV\/-\/type hierarchies of symmetries for~$\cE$. Having thus reduced the problem to the Hamiltonian case, we
calculate the Lie\/-\/type brackets, transferred from the commutators of the symmetries in the images of the operators~$\square$ onto their domains. With all this, we describe the generators and derive all the commutation relations in the symmetry algebras of the 2D~Toda chains, which serve here as an illustration for a much more general algebraic and geometric set\/-\/up.
\end{abstract}
\maketitle

\subsection*{Introduction}
In the paper~\cite{Twente}, 
we introduced a well\/-\/defined notion of linear matrix operators in total derivatives, whose images in the Lie algebras of evolutionary vector fields on the jet spaces are closed with respect to the commutation. This yields a generalization for the classical theory of recursion operators and Poisson structures for integrable systems. We explained how each operator transfers the commutation of the vector fields to the Lie brackets with bi-differential structural constants on the quotient of its domain by the kernel.

Second, we associated such operators with the 2D~Toda chains
\begin{equation}\label{IEToda}
\cEToda=\Bigl\{u^i_{xy}=\exp\bigl(\sum_{j=1}^m K^i_{\,j}u^j\bigr),
1\leq i\leq m\Bigr\}
\end{equation}
related to semi\/-\/simple complex Lie algebras~\cite{Leznov,LeznovSmirnovShabat}.
We derived an explicit formula for the operators that factor higher symmetries of these chains. Using the auxiliary matrix operators that we proved to be Hamiltonian, we elaborated a procedure that yields all the commutation relations in the symmetry Lie algebras~$\sym\cEToda$ 
(naturally, these symmetry algebras are not commutative). This solved a 
long\/-\/standing problem in geometry of differential equations and completed previously known results by Leznov, Meshkov, Shabat, Sokolov, and others 
(see~\cite{LeznovSmirnovShabat,Meshkov198x,ShabatYamilov,SokolovUMN}
and references therein).

Actually, the general scheme of~\cite{Twente} is applicable, in particular, for the description of symmetry algebras for a wider class of the Euler\/--\/Lagrange hyperbolic systems of Liouville type~\cite{ShabatYamilov,SokolovUMN}. Moreover, the group analysis of integrable systems, as a motivation, results in the well\/-\/defined concept of operators whose images span involutive distributions on the jet spaces, but not on differential equations, which is of an independent interest.

In this note, we illustrate the reasonings of~\cite{Twente} using the root systems of the complex semi\/-\/simple Lie algebras of rank two.
Among all two\/-\/component exponential nonlinear systems~\eqref{IEToda}, these 2D~Toda chains with $K$~Cartan matrices admit the largest groups of conservation laws~\cite{ShabatYamilov} and are integrable in quadratures~\cite{Leznov}.

The equality of the rank to two means the following:
\begin{itemize}
\item
The hyperbolic Toda chains~\eqref{IEToda} upon $u^1$, $u^2$ are,
we repeat, \emph{two}\/-\/component.
\item
The number of vector fields~$Y_i$ that generate the characteristic Lie
algebra through commutators 
(see section~\ref{SecA2ChAlg} below and~\cite{LeznovSmirnovShabat}) 
equals \emph{two}. Also, the numbers of linear independent iterated commutators 
$Y_{(i_1,\ldots,i_k)}=[Y_{i_1}$,$[\ldots[Y_{i_{k-1}}$,$Y_{i_k}]\ldots]]$ fall at most \emph{twice}, and the accumulated sum of these dimension's falls equals \emph{two}. Thence, by the Frobenius theorem, 
\emph{two} invariants~$w^1$,\ $w^2$ appear.%
\footnote{For example, the paper~\cite{Murtazina} contains a brute force classification of integrable one\/-\/component hyperbolic equations 
with respect to the low\/-\/dimensional characteristic Lie algebras.}
Using the characteristic Lie algebras, we introduce \emph{two} finite sequences of the adapted coordinates, which simplifies the description of these invariants. On the other hand, we use the \emph{two} invariants for replacing the derivatives of the \emph{two} dependent variables at all sufficiently high differential orders.
\item
Conservation laws for Toda system~\eqref{IEToda} are differentially generated
(up to $x\leftrightarrow y$)
by these \emph{two} invariants, which are solutions of the characteristic equation $D_y(w)\doteq0$ on~$\cEToda$. 
\item
Higher symmetries of the Toda chain~\eqref{IEToda} have a functional freedom and are parameterized by \emph{two} functions $\phi^1$,\ $\phi^2$ that depend on $x$ and any derivatives of the integrals $w^i$ up to a certain differential order.
\item
The differential operators~$\square$ that yield symmetries of~\eqref{IEToda}, when applied to the tuples $\bigl(\phi^1,\phi^2\bigr)$, are $(2\times2)$-\/matrices.
\item
The Lie algebra structures transferred from $\sym\cEToda$ to the domains of~$\square$ are described by the bi\/-\/differential brackets $\ib{\,}{\,}{\square}$ that contain \emph{two} components.
\item
The Hamiltonian structures~$\smash{\hat{A}_k}$ 
that are defined on the domains of the operators~$\square$ but take values elsewhere (in the Lie algebra of velocities of the integrals~$w^i$,
see~\cite{Twente}), are also $(2\times2)$-\/matrices. Likewise, the brackets $\ib{\,}{\,}{\smash{\hat{A}_k}}$ transferred onto the domains of~$\smash{\hat{A}_k}$ from the commutators of Hamiltonian vector fields in their images are also \emph{two}\/-\/component (thence the equality $\ib{\,}{\,}{\smash{\hat{A}_k}}=\ib{\,}{\,}{\square}$ makes sense).
\item
The KdV\/-\/type hierarchies of velocities of $w^i$ and the modified KdV\/-\/type hierarchies of commuting Noether symmetries of~$\cEToda$, which are related by \emph{two}\/-\/component Miura's substitutions with the former, are composed by the (right\/-\/hand sides of) \emph{two}-\/component evolutionary systems.
\end{itemize}

All constructions and notation follow~\cite{Twente} except for the characteristic Lie algebras that were introduced in~\cite{LeznovSmirnovShabat} and were discussed in detail in~\cite{Shabat95}.
All notions from the geometry of PDE are standard (see~\cite{ClassSym,Opava,Olver}) and have been surveyed in~\cite[sect.~2]{Twente}.
All extensive calculations were performed using the software~\cite{Jets}.

To start with, we recall that in the fundamental paper~\cite{ShabatYamilov},
A.~B.~Shabat \textit{et al.} proved the existence of maximal ($r=\bar{r}=m$)
sets~\eqref{IInt} of conserved densities 
\begin{equation}\label{IInt}
w_1,\ldots,w_r\in\ker D_y\bigr|_{\cEToda},\qquad
\bar{w}_1,\ldots,\bar{w}_{\bar{r}}\in\ker D_x\bigr|_{\cEToda}.
\end{equation}
for~$\cEToda$ if and only if the matrix~$K$ in~\eqref{IEToda} 
is the Cartan matrix of a root system
for a semi\/-\/simple complex Lie algebra of rank~$r$, which is always the case in what follows with~$r=2$.
Note that the integrals~\eqref{IInt} allow to replace the derivatives of unknown functions of any sufficiently high order using the derivatives of the integrals. In~\cite{Shabat95}, A.~B.~Shabat proposed an iterative procedure that specifies an adapted system of the remaining lower\/-\/order coordinates
and that makes linear the \emph{coefficients} of the linear first\/-\/order characteristic equation $D_y(w)\doteq0$ on~$\cEToda$. That algorithm is self\/-\/starting, simplifies considerably the search for the first integrals of the characteristic equation, and gives the estimate for the differential orders of solutions.

Another method (which we do not use here) for finding the first integrals is based on the use of Laplace's invariants
, see~\cite{SokolovUMN}. 
The authors of that paper investigated (primarily, in the case of one unknown function and one equation~$\cE$ upon it)
the operators that factor symmetries of~$\cE$. Also there, the pioneering idea to study the operators whose images are closed under the commutation was proposed. We indicate further 
the papers~\cite{Demskoi2004,TMPhGallipoli,SokStar,Startsev2006} that address the problem of construction of such factoring operators for multi\/-\/component hyperbolic systems of the Liouville type. 

The general concept of operators whose images determine involutive distributions, the definition itself and the classification, has been elaborated in~\cite{Twente}. There, as a by\/-\/product, we obtained an explicit formula for the operators~$\square$ that factor higher symmetries of the Euler\/--\/Lagrange Liouville\/-\/type systems and for the bi\/-\/differential brackets on their domains. The former yields all the generators of the symmetry algebras for such systems, and the latter describes all the commutation relations.

\smallskip%
This paper is structured as follows.
First we outline the basic concept using the scalar Liouville equation $u_{xy}=\exp(2u)$ as the motivating example. This covers the case of the root system~$\mathsf{A}_1$. 
The following fact, which holds true for any rank $r\geq1$, 
is very convenient in practice
: the differential orders of the $r$~integrals $w^1$,\ $\ldots$,\ $w^r$ for the 2D~Toda systems~\eqref{IEToda} associated with the complex semi\/-\/simple Lie algebras~$\gothg$ are equal (up to a shift by $+1$) to the gradations for the principal realizations of the basic
representations of the respective untwisted affine Lie algebras~$\gothg^{(1)}$
(see, \textit{e.g.}, the \emph{list} in~\cite[\S14.2]{Kac}).

Then we realize the geometric scheme
\[
D_y(w)\doteq0\ \longmapsto\ w^1,w^2\ \longmapsto\ \square
\ \longmapsto\ \smash{\hat{A}_k}\ \longmapsto\ \ib{\,}{\,}{\square};\qquad
\hat{A}_k\ \longmapsto\ \hat{A}_1^{(\cdot)},
\]
for the simple complex rank two Lie algebras (for the root systems
$\mathsf{A}_2$, $\mathsf{B}_2\simeq\mathsf{C}_2$, and~$\mathsf{G}_2$).
In other words, we associate the operators to invariants, the brackets
to operators, and find the deformations of the Poisson structures.
Only once, for the root system~$\mathsf{A}_2$, we calculate the characteristic Lie algebra for the corresponding 2D~Toda chain~\eqref{IEToda} and obtain the integrals~$w^1,w^2$ using the adapted system of coordinates. We modify the scheme of~\cite{Shabat95} such that, first, the two\/-\/component Toda chain is not represented as a reduction of the infinite chain and, second, we do not introduce an excessive third field which is compensated by a constraint (as in~\cite{Shabat95}).

\begin{rem}
We do not of course re\/-\/derive the structures for the algebra~$\mathsf{D}_2=\mathsf{A}_1\oplus\mathsf{A}_1$ that is not simple, since the operator~$\square$ and KdV's second Hamiltonian structure~$\smash{\hat{A}_k}$ are known for each of the two uncoupled components of the chain~\eqref{IEToda} with $K=\left(\begin{smallmatrix}2&0\\ 0&2\end{smallmatrix}\right)$, see Example~\ref{ExLiou} below. However, the ``$x$-\/component'' of the full symmetry algebra with the generators
\[
\vph=\begin{pmatrix}\square^{\#1} & 0\\ 0 & \square^{\#2}\end{pmatrix}
\binom{\phi^1\bigl(x,[w^{\#1}],[w^{\#2}]\bigr)}%
      {\phi^2\bigl(x,[w^{\#1}],[w^{\#2}]\bigr)}
\]
is \emph{not} just the direct sum of the two symmetry subalgebras for the two Liouville equations. Indeed, this formula shows that the integrals can be intertwined in the generators, although the fields are not coupled in 
system~\eqref{IEToda} with the choice of~$K$ as above, and it
proves that the symmetries intertwine the fields.
\end{rem}

\begin{rem}
Each of the operators~$\square$, which we obtain from the $r$ integrals, consists of~$r$ columns, one column for each integral. In the $r=2$ case, only the respective first columns were specified in the encyclopaedia~\cite{Ibragimov}, 
see also~\cite{Meshkov198x}. Here, we complete the description of the
symmetry generators.
\end{rem}

\section{Basic concept}
Let us begin with a motivating example.
\begin{example}[The Liouville equation]\label{ExLiou}
Consider the scalar Liouville equation
\begin{equation}\label{ELiou}
\cELiou=\{u_{xy}=\exp(2u)\}.
\end{equation}
The differential generators $w$, $\bar{w}$ of its conservation laws
$[\eta]=\int f(x,[w])\,\Id x+\int \bar{f}(y,[\bar{w}])\,\Id y$ are
\begin{equation}\label{KdVSubst}
w=u_x^2-u_{xx} \quad\text{and}\quad \bar{w}=u_y^2-u_{yy}
\end{equation}
such that $D_y(w)\doteq0$ and $D_x(\bar{w})\doteq0$ by virtue ($\doteq$)
of~$\cELiou$ and its differential consequences. The operators
\begin{equation}\label{SquareSok}
\square=u_x+\tfrac{1}{2}D_x\quad\text{and}\quad \bar{\square}=u_y+\tfrac{1}{2}D_y
\end{equation}
factor higher and Noether's symmetries
\[
\vph=\square\bigl(\phi(x,[w])\bigr),\quad
\vph_\cL=\square\Bigl(\frac{\delta\cH(x,[w])}{\delta w}\Bigr); \qquad
\bar{\vph}=\bar{\square}\bigl(\bar{\phi}(y,[\bar{w}])\bigr),\quad
\bar{\vph}_\cL=\bar{\square}\Bigl(\frac{\delta\bar{\cH}(y,[\bar{w}])}{\delta\bar{w}}\Bigr)
\]
of the Euler\/--\/Lagrange equation~\eqref{ELiou}
for any smooth $\phi,\bar{\phi}$ and $\cH,\bar{\cH}$.
Note that the operator $\square=\tfrac{1}{2}D_x^{-1}\circ
\bigl(\ell_w^{(u)}\bigr)^*$ is obtained using the adjoint linearization
of~$w$, and similarly for~$\bar{\square}$.

Each of the images of~\eqref{SquareSok} is closed w.r.t.\ the commutation such that
\[
\bigl[\square(p),\square(q)\bigr]=\square\bigl([p,q]_\square\bigr),
\]
where the bracket $[\,,\,]_\square$ on the domain of~$\square$ admits the
standard decomposition (the vector field 
$\cEv_\vph=\sum_k D_x^k(\vph)\cdot\dd/\dd u_k$
is the evolutionary derivation),
\[
[p,q]_\square=
  \cEv_{\square(p)}(q)-\cEv_{\square(q)}(p)+\ib{p}{q}{\square}.
\]
For the operator~$\square$ on the Liouville equation, the bi\/-\/differential bracket $\ib{\,}{\,}{\square}$ is
\[
\ib{p}{q}{\square}=D_x(p)\cdot q-p\cdot D_x(q),
\]
and similar formulas hold for the operator~$\bar{\square}$.
The symmetry algebra $\sym\cELiou\simeq\img\square+\img\bar{\square}$ is the sum of images of~\eqref{SquareSok},
and the two summands commute between each other,
$  
[\img\square,\img\bar{\square}]\doteq0
$ on~$\cELiou$. 
Therefore,
\begin{equation}\label{IStrongSquare}
[\img\square+\img\bar{\square},\img\square+\img\bar{\square}]\subseteq
\img\square+\img\bar{\square}.
\end{equation}

The 
operator~$\square$ factors higher symmetries
of the potential modified KdV equation
\begin{equation}\label{IpmKdV}
\cE_{\text{pmKdV}}=\{u_t=-\tfrac{1}{2}u_{xxx}+u_x^3=\square(w)\},
\end{equation}
whose commutative hierarchy is composed by
Noether's symmetries $\vph_\cL\in\img(\square\circ\delta/\delta w)$ of the Liouville equation~\eqref{ELiou}.
The operator~$\square$ factors the second Hamiltonian structure $B_2=\square\circ A_1\circ\square^*$ for~$\cE_{\text{pmKdV}}$, 
here $A_1=D_x^{-1}=\hat{A}_1^{-1}$ is the first Hamiltonian operator for the
potential KdV equation and equals the inverse of the first Hamiltonian operator
for~KdV.

The generator~$w$ of conservation laws for~$\cELiou$ provides the Miura substitution~\eqref{KdVSubst}
from $\cE_{\text{pmKdV}}$ to the Korteweg\/--\/de Vries equation
\begin{equation}\label{IKdV}
\cE_{\text{KdV}}=\{w_t=-\tfrac{1}{2}w_{xxx}+3ww_x\}.
\end{equation}
The second Hamiltonian structure for~$\cE_{\text{KdV}}$ is factored to
the product $\hat{A}_2=\square^*\circ\hat{B}_1\circ\square$, where
$\hat{B}_1=D_x$ is the first Hamiltonian structure for the modified~KdV.
%
%
The bracket $\ib{\,}{\,}{\square}$ on the domain of~$\square$ is equal to the
bracket $\ib{\,}{\,}{\hat{A}_2}$ induced on the domain of the
operator~$\smash{\hat{A}_2}$ (which is Hamiltonian and hence its image is
closed under commutation) for~$\cE_{\text{KdV}}$.
In what follows, we refer to these correlations as standard, see~\cite{Twente}.
\end{example}

\begin{define}[\cite{SokolovUMN}]\label{DefLiouType}
A \emph{Liouville\/-\/type system}%
\footnote{There exist other, non\/-\/equivalent definitions of 
the Liouville type systems.}
$\cEL$ is a system $\{u_{xy}=F(u,u_x,u_y;x,y)\}$ of hyperbolic equations 
which admits nontrivial \emph{first integrals} 
\[
w_1,\ \ldots,\ w_r\in\ker D_y{\bigr|}_{\cEL};\quad 
\bar{w}_1,\ \ldots,\ \bar{w}_{\bar{r}}\in\ker D_x{\bigr|}_{\cEL}
\]   
for the linear first order characteristic equations
${D_y\bigr|}_{\cEL}(w_i)\doteq0$ and ${D_x\bigr|}_{\cEL}(\bar{w}_j)\doteq0$ 
that hold by virtue ($\doteq$) of~$\cEL$, and such that
all conservation laws for $\cEL$ are of the form
$\int f(x,[w])\,\Id x\oplus\int g(y,[\bar{w}])\,\Id y$.
\end{define}     

\begin{example}
The $m$-\/component 2D~Toda chains~\eqref{IEToda} associated with semi\/-\/simple
complex Lie algebras~\cite{Leznov} constitute an important class
of Liouville\/-\/type systems, here $u=(u^1,\ldots,u^m)$.
Further on, we consider these exactly solvable systems, bearing in mind that the reasonings remain applicable to a wider class of the Euler\/--\/Lagrange Liouville\/-\/type systems~$\cEL$.
\end{example}

\begin{rem}
The 2D~Toda systems~\eqref{IEToda} are 
Euler\/--\/Lagrange, with the Lagrangian density
$L=-\tfrac{1}{2}\langle\kappa u_x,u_y\rangle-\mathrm{H}_{\IL}(u;x,y)$.
The $(m\times m)$-\/matrix~$\kappa$ with the entries
\[
\kappa_{ij}=\frac{2\langle\alpha_i,\alpha_j\rangle}{|\alpha_i|^2\cdot
   |\alpha_j|^2}=\frac{1}{|\alpha_i|^2}\cdot K^i_{\,j}
\]
is determined by the simple roots~$\alpha_k$ of the semi\/-\/simple Lie algebra.

Let $\gm=\dd L/\dd u_y$ be the momenta, then it can be readily seen that
the integrals $w^1$,\ $\ldots$,\ $w^m$ of the characteristic equation 
are differential functions $w^i=w^i[\gm]$ in~$\gm$.
\end{rem}

\begin{state}
The differential orders of the integrals~$w^i$ with respect to 
the momenta~$\gm$ for the 2D~Toda chains~\eqref{IEToda} associated with complex semi\/-\/simple Lie algebras~$\gothg$ coincide with the gradations 
for the principal realizations of the basic 
(\textit{i.e.}, simplest nontrivial highest weight, see~\cite{Kac})
representations 
of the corresponding untwisted affine Lie algebras~$\gothg^{(1)}$.
\end{state}

The integrals $w^i$ for a nonlinear Liouville\/-\/type hyperbolic system can be obtained using an iterative procedure that is illustrated 
in section~\ref{SecA2ChAlg} below. In the meantime, we assume that 
the integrals are already known. Let them be \emph{minimal}, meaning that
$f\in\ker D_y\bigr|_{\cEL}$ implies $f=f(x,[w])$.

\begin{theorNo}[\cite{Twente}]
Let the above assumptions and notation hold.
Introduce the operator
\begin{equation}\label{ISquare}
\square=\bigl(\ell_w^{(\gm)}\bigr)^*,
\end{equation}
which is the operator adjoint to the linearization (the Fre\-ch\'et derivative) of the integrals~$w$ w.r.t.\ the momenta~$\gm$.
Then we claim the following:
{\renewcommand{\theenumi}{\roman{enumi}}\begin{enumerate}
\item
All (up to $x\leftrightarrow y$) Noether symmetries $\vph_\cL$ of the
Lagrangial~$\cL$ for~$\cEL$ are
\[
\vph_\cL=\square\Bigl(\frac{\delta\cH(x,[w])}{\delta w}\Bigr)\qquad
\text{for any $\cH$.}
\]
\item
All (up to $x\leftrightarrow y$) symmetries $\vph$ of the system~$\cEL$ are
\[
\vph=\square\bigl(\phi(x,[w])\bigr)\qquad\text{for any $\phi=(\phi^1,\ldots,
  \phi^r)$. 
}
\]
\item
In the chosen system of coordinates, the image of the operator~$\square$ is
closed with respect to the commutation in the Lie algebra~$\sym\cEL$.
\item
Under a diffeomorphism $\tilde{w}=\tilde{w}[w]$, the $r$-\/tuples~$\phi$ are
transformed by
\[
\phi\mapsto\tilde{\phi}=
   \bigl[\bigl(\ell_{\tilde{w}}^{(w)}\bigr)^*\bigr]^{-1}(\phi).
\]
Therefore, under any reparametrization $\tilde{u}=\tilde{u}[u]$ of
the dependent variables $\vec{u}={}^t(u^1,\ldots,u^m)$ in equation~$\cEL$,
and under a simultaneous change $\tilde{w}=\tilde{w}[w]$, the
operator~$\square$ obeys the transformation rule
\[
\square\mapsto\tilde{\square}=\ell_{\tilde{u}}^{(u)}\circ \square\circ
 \bigl(\ell_{\tilde{w}}^{(w)}\bigr)^*
  \Bigr|_{\substack{w=w[u]\\u=u[\tilde{u}]}}.
\]
Consequently, the operator~$\square$ 
becomes well defined:
it is a \emph{Frobenius} operator of \emph{second kind},
see~\cite{Twente}. 
\item
The operator
\begin{equation}\label{Quattro}
\smash{\hat{A}_k}=\square^*\circ\bigl(\ell_\gm^{(u)}\bigr)^*\circ\square
\end{equation}
is Hamiltonian. 
\item
The bracket $\ib{\,}{\,}{\square}$ on the domain 
of the operator~$\square$ satisfies the equality
\begin{equation}\label{BothBrackets}
\ib{\,}{\,}{\square}=\ib{\,}{\,}{\smash{\hat{A}_k}}.
\end{equation}
Its right\/-\/hand side is calculated explicitly by using 
the formula (\cite{Opava,Olver}, see also~\cite{Twente}) 
that is valid for Hamiltonian operators~$\smash{\hat{A}_k}=\|\sum\limits_\tau A^{\alpha\beta}_\tau\cdot D_\tau\|$,
\begin{equation}\label{EqDogma}
\ib{p}{q}{\hat{A}_k}^i=\sum_{\sigma,\alpha} (-1)^\sigma
 \Bigl(D_\sigma\circ\Bigl[\sum_{\tau,\beta} D_\tau(p^\beta)\cdot
 \frac{\dd A_\tau^{\alpha\beta}}{\dd u^i_\sigma}\Bigr]\Bigr)
 \bigl(q^\alpha\bigr).
\end{equation}
This yields the commutation relations in the Lie algebra~$\sym\cEL$.
\item\label{DependOnW}
All coefficients of the operator~$\smash{\hat{A}_k}$ and of the
bracket~$\ib{\,}{\,}{\square}$ are differential functions
of the minimal conserved densities~$w$ for~$\cEL$.
\end{enumerate}}
\end{theorNo}

\noindent%
The above theorem is our main instrument that describes 
all the symmetry generators for 2D~Toda chains and calculates 
all the commutation relations in the symmetry algebras.

\section{The root system $\mathsf{A}_2$}
Consider the Euler\/--\/Lagrange 2D~Toda system associated with the simple Lie
algebra~$\mathfrak{sl}_3(\BBC)$, see~\cite{Leznov,LeznovSmirnovShabat,Shabat95},
\begin{equation}\label{A2Toda}
\cEToda=\Bigl\{ u_{xy}=\exp(2u-v),\ v_{xy}=\exp(-u+2v),\qquad
K=\left(\begin{smallmatrix}\phantom{+}2&-1\\
 -1&\phantom{+}2\end{smallmatrix}\right)\Bigr\}.
\end{equation}

\subsection{The characteristic Lie algebra}\label{SecA2ChAlg}
First 
we realize two itegrations of the self\/-\/adaptive method 
from~\cite{Shabat95}, which is based on the use of the characteristic Lie algebra, and we obtain two integrals $w^1$,\ $w^2$ of the characteristic
equation $D_y(w)\doteq0$ on~\eqref{A2Toda}.
Our reasonings differ from the original approach of~\cite{Shabat95}: we do not introduce excessive dependent variables and hence do not need to compensate their presence with auxiliary constraints.

\smallskip%
Our remote goal is a choice of three layers of the adapted variables 
$b_0^1$,\ $b_0^2$,\ $b_1^1$,\ $b_1^2$, and~$b_2^1$ such that all the coefficients of the linear characteristic equation also become linear.
Then all the integrals will be found easily, expressed in these variables.
The number of the adapted variables is specified by the problem,
and we have to confess that, actually, $b_1^2$~will be redundant \textit{a posteriori} because it will be replaced using the integral~$w^1$ in the end.

\subsubsection*{Step 1}
Regarding the exponential functions
\[
c(i)\mathrel{{:}{=}}\exp\Bigl(\sum\nolimits_j K^i_{\,j}u^j\Bigr)
\]
in the right\/-\/hand sides of the Toda equations~\eqref{IEToda} as linear independent, collect the coefficients $Y_i$ of $c(i)$ in the total derivative
\[
D_y=\sum_{i=1}^m c(i)\cdot Y_i.
\]
Clearly, the solution of the characteristic equation $D_y(w)\doteq0$ on the Toda chain is equivalent to solution of the system $\bigl\{Y_i(w)=0$, 
$1\leq i\leq m\bigr\}$.

For system~\eqref{A2Toda}, we obtain the vector fields
\begin{equation}\label{YiStep1}
\begin{aligned}
Y_1&=\frac{\dd}{\dd u_x}+(2u_x-v_x)\,\frac{\dd}{\dd u_{xx}}+
   \bigl(\underline{(2u_x-v_x)^2}
    +(2u_{xx}-v_{xx})\bigr)\,\frac{\dd}{\dd u_{xxx}}+\cdots,\\
Y_2&=\frac{\dd}{\dd v_x}+(2v_x-u_x)\,\frac{\dd}{\dd v_{xx}}+
   \bigl(\underline{(2v_x-u_x)^2}
    +(2v_{xx}-u_{xx})\bigr)\,\frac{\dd}{\dd v_{xxx}}+\cdots.
\end{aligned}
\end{equation}
The underlined terms are \emph{quadratic} in derivatives of the fields, 
and it is our task to make them linear by introducing a convenient system of local coordinates (see take~2 of step~3 below).

Taking the iterated commutators
\[
Y_{(i_1,\ldots,i_k)}\mathrel{{:}{=}}
   \bigl[Y_{i_1},[\ldots,[Y_{i_{k-1}},Y_{i_k}]\ldots]\bigr]
\]
of the basic vector fields~$Y_i$, we generate the \emph{characteristic Lie algebra}~\cite{LeznovSmirnovShabat,Shabat95} for the Toda chain. If this algebra is finite dimensional (which is the case here), then the exponential\/-\/nonlinear system~\eqref{IEToda} is exactly solvable in quadratures; if the characteristic algebra admits a finite dimensional representation, system~\eqref{IEToda} is integrable by the inverse scattering (\textit{ibid}).
For any root system and the Chevalley generators $\mathfrak{e}_n$,\ $\mathfrak{f}_n$, and~$\mathfrak{h}_n$ of the semi\/-\/simple Lie algebra~$\gothg$,
see~\cite{Humphreys}, the characteristic Lie algebra is isomorphic to the Lie subalgebra of~$\gothg$ generated by the Chevalley generators~$\mathfrak{f}_n$, see~\cite{LeznovSmirnovShabat}.

For~$\mathsf{A}_2$, we obtain the commutator
\[
Y_{(2,1)}=-\frac{\dd}{\dd u_{xx}}+\frac{\dd}{\dd v_{xx}}
  -3u_x\,\frac{\dd}{\dd u_{xxx}}+3v_x\,\frac{\dd}{\dd v_{xxx}}+\cdots.
\]
(This manifests a general fact that is always true: the leading terms of 
the $(k+1)$-st iterated commutators are the derivations w.r.t. 
some derivatives $u^i_{k+1}$, whose order is higher than in the leading terms of the preceding, $k$-th, iterated commutators.)
We finally note that all the triple commutators, $Y_{(1,2,1)}$ and $Y_{(2,2,1)}$, vanish.

By the Frobenius theorem, a fall of the number of linear independent iterated commutators at the $i$th step is equal to the number of first integrals of the characteristic equation that appear at this step. The differential order of these new integrals for the Toda chains will be~$i+1$.

For the system~\eqref{A2Toda}, there appears one ($1=\dim\langle Y_i\rangle-
\dim\langle Y_{(i_1,i_2)}\rangle$) integral, $w^1$, of order~$2$. 
The second and last one ($1=\dim\langle Y_{(i_1,i_2)}\rangle - 
\langle Y_{(i_1,i_2,i_3)}\equiv0\rangle$), the integral~$w^2$, has order~$3$.
For arbitrary root systems, the differential orders (shifted by $+1$) of the integrals are described by the proposition in the previous section.

\subsubsection*{Step 2}
Our remote goal, see above, will be achieved when the expansion
\[
D_x=\sum_{i=1}^m b_1^i\,Y_i + \sum_{i=1}^{m-1} b_2^i\,Y_{(i+1,i)}+\cdots
\mod \mathfrak{Z}\colon\ker D_y{\bigr|}_{\cEL}\to\ker D_y{\bigr|}_{\cEL}
\]
is found for the \emph{other} total derivative,~$D_x$.
Here the vector field~$\mathfrak{Z}$ contains only the derivations w.r.t.\ the integrals (yet unknown) and their derivatives, and the dots stand for finitely many summands provided that the characteristic algebra is finite dimensional.
The former idea exprimes the replacement of the higher order field derivatives, $u^i_k$ with $k\gg1$, using the integrals, while the latter assumption is again based on the fact that there are as many integrals as the fields for $K$~Cartan matrices.

By definition, put $b_0^i\mathrel{{:}{=}}u_x^i$.

Substituting the vector fields~$Y_i$ contained in~\eqref{YiStep1} for
$\dd/\dd u_x^i$ in~$D_x$, we obtain the expansion
\[
D_x=u_{xx}\,Y_1+v_{xx}\,Y_2+\cdots,
\]
where the dots stand for the derivations w.r.t.\ second and higher order derivatives of the dependent variables. Consequently, we set
\[
b_1^i\mathrel{{:}{=}}u^i_{xx}.
\]

\subsubsection*{Step 3}
Using the four adapted coordinates $b_0^i$ and~$b_1^j$, we rewrite the basic vector fields~$Y_k$ as follows,
\[
Y_1=\frac{\dd}{\dd b_0^1}+(2b_0^1-b_0^2)\,\frac{\dd}{\dd b_1^1}+\cdots,\qquad
Y_2=\frac{\dd}{\dd b_0^2}+(-b_0^1+2b_0^2)\,\frac{\dd}{\dd b_2^1}+\cdots.
\]
Solving now the system
\[
Y_1(w^1)=0,\ Y_2(w^1)=0\qquad\text{for $w^1(b_0^1,b_0^2,b_1^1,b_1^2)$,}
\]
we obtain the integral
\[
w^1=u_{xx}+v_{xx}-u_x^2+u_x v_x-v_x^2.
\]
Note that, from now on, the coordinate $v_{xx}$ and its descendants can be replaced using $w^1$, $u_{xx}$, and first order derivatives.

\subsubsection*{Step 1, take 2}
Within the second iteration of the algorithm, we repeat steps~1--3 advancing one term farther in the expansions.

Let us indeed replace $v_{xx}$ (although it remains an adapted coordinate)
with~$w^1$. Therefore we expand the vector field~$D_y$ as
\[
D_y=c(1)\cdot Y_1+c(2)\cdot\frac{\dd}{\dd v_x} \mod\mathfrak{Z}\colon
   \ker D_y{\bigr|}_{\cEL}\to 0,
\]
where the derivations w.r.t.\ $w^1$ and its descendants cut off 
the `$v$-\/part' of the total derivative~$D_y$.
This yields
\[
Y_{(2,1)}=-\frac{\dd}{\dd u_{xx}}+\cdots,
\]
but now the commutator does not contain any derivations w.r.t.\ the derivatives of~$v$. 

\subsubsection*{Step 2, take 2}
Using the three vector fields, $Y_1$, $Y_2$, and~$Y_{(2,1)}$, we rewrite
\[
D_x=u_{xx}\,Y_1+v_{xx}\,Y_2
   +\bigl((2u_x-v_x)u_{xx}-u_{xxx}\bigr)\cdot Y_{(2,1)}+\cdots
   \mod\mathfrak{Z}\colon\ker D_y{\bigr|}_{\cEL}\to\ker D_y{\bigr|}_{\cEL}.
\]
Consequently, we set
\[
b_2^1\mathrel{{:}{=}}(2u_x-v_x)u_{xx}-u_{xxx}.
\]

\subsubsection*{Step 3, take 2}
Calculating the derivative $D_y{\bigr|}_{\cEL}(b_2^1)$, we substitute it 
in~$D_y$ and collect the coefficients $Y_i$ of the exponential 
nonlinearities~$c(i)$ in this total derivative.

The result is beyond all hopes: the coefficients of both fields, 
$Y_1$ and~$Y_2$, are \emph{linear} in the adapted coordinates,
\[
\begin{aligned}
Y_1&=\frac{\dd}{\dd b_0^1}+\bigl(2b_0^1-b_0^2\bigr)\cdot\frac{\dd}{\dd b_1^1}
   +b_1^2\,\frac{\dd}{\dd b_2^1}+\cdots,\\
Y_2&=\frac{\dd}{\dd b_0^2}+\bigl(-b_0^1+2b_0^2\bigr)\cdot\frac{\dd}{\dd b_1^2}
   -b_1^1\,\frac{\dd}{\dd b_2^1}+\cdots.
\end{aligned}
\]
In other words, the quadratic terms, which were underlined in~\eqref{YiStep1},
are transformed into the linear ones. This is due to the quadratic nonlinearity
in the new adapted variable~$b_2^1$.

Finally, we solve the characteristic equation
\[
Y_1(w^2)=0,\ Y_2(w^1)=0\qquad\text{for $w^2(b_0^1,b_0^2,b_1^1,b_1^2,b_2^1)$}
\]
under the assumption\footnote{A practically convenient feature of the algorithm is that it allows to fix the `top' (the higher order terms) of the first integrals in advance, whence the redundant freedom in adding derivatives of the previously found lower order solutions is eliminated.}
$\dd w^2/\dd b_2^1\neq0$.
We find the solution
\[
w^2=-b_2^1-b_0^2b_1^1+b_0^1b_1^2+{(b_0^1)}^2 b_0^2-b_0^1{(b_0^2)}^2.
\]
Returning to the original notation, we obtain
\[
w^2=u_{xxx}-2u_xu_{xx}+u_xv_{xx}+u_x^2v_x-u_xv_x^2.
\]
Obviously, the integral~$w^2$ can be used to replace the derivative $u_{xxx}$
and its differential consequences. 

We conclude that now, at the endpoint of the algorithm, both total derivatives, $D_x$ and~$D_y$, 
contain finitely many terms modulo the vector fields 
that preserve (respectively, annihilate) the kernel~$\ker D_y{\bigr|}_{\cEL}$.

In what follows, we do not repeat similar iterative reasonings for 
the root systems $\mathsf{B}_2$ (see~\eqref{B2Int}) and $\mathsf{G}_2$
(see p.~\pageref{G2Int}), but write down at once the integrals of orders $2$,~$4$ and $2$,~$6$, respectively. The second integral~$w^2$ for $\mathsf{B}_2$ (with a minor misprint in the last term) and the higher order `top' for~$w^2$ 
for~$\mathsf{G}_2$ are available in the encyclopaedia~\cite{Ibragimov}.

\subsection{The symmetry algebra: operators and brackets}\label{SecA2Sym}
From the previous section, we know the minimal integrals,
\begin{align*}  
w^1&=u_{xx}+v_{xx}-u_x^2+u_xv_x-v_x^2,\\ 
w^2&=u_{xxx}-2u_xu_{xx}+u_xv_{xx}+u_x^2v_x-u_xv_x^2,
\end{align*}
for the 2D~Toda chain~\eqref{A2Toda} associated with 
the root system~$\mathsf{A}_2$. Hence we are at the starting point
for describing its symmetry algebra and revealing the corresponding Poisson structures and the KdV\/-\/type hierarchies.

Let us introduce the momenta
\[
\gm^1\mathrel{{:}{=}}2u_x-v_x,\qquad \gm^2\mathrel{{:}{=}}2v_x-u_x,
\]
whence we express the integrals as follows,
\begin{align*}
w^1 &= 3\gm^1_x+3\gm^2_x-(\gm^1)^2-\gm^1\gm^2-(\gm^2)^2,\\
w^2 &= 2\gm^1_{xx}+\gm^2_{xx}-2\gm^1\gm^1_x-\gm^2\gm^1_x+\tfrac{2}{9}(\gm^1)^3
+\tfrac{1}{3}(\gm^1)^2\gm^2-\tfrac{1}{3}\gm^1(\gm^2)^2-\tfrac{2}{9}(\gm^2)^3.
\end{align*}
By the general scheme of~\cite{Twente},
all symmetries (
up to $x\leftrightarrow y$) of~\eqref{A2Toda} are of the form
$\vph=\square\bigl(\vec{\phi}\bigl(x,[w^1],[w^2]\bigr)\bigr)$, where
$\vec{\phi}={}^t(\phi^1,\phi^2)$ is a pair of arbitrary functions 
and the $(2\times2)$-\/matrix operator in total derivatives is
given by formula~\eqref{ISquare},
\begin{equation}\label{SquareA2}
\square=\begin{pmatrix}
u_x+D_x & 
 -\tfrac{2}{3}D_x^2-u_xD_x-\tfrac{1}{3}u_x^2-\tfrac{2}{3}u_xv_x
   +\tfrac{2}{3}v_x^2+\tfrac{1}{3}u_{xx}-\tfrac{2}{3}v_{xx} \\
v_x+D_x & 
 -\tfrac{1}{3}D_x^2+\tfrac{2}{3}u_{xx}-\tfrac{1}{3}v_{xx}
   -\tfrac{2}{3}u_x^2+\tfrac{2}{3}u_xv_x+\tfrac{1}{3}v_x^2 \end{pmatrix}.
\end{equation}
Next, we calculate the Hamiltonian operator~\eqref{Quattro},
\[
\smash{\hat{A}_k}=\begin{pmatrix} A_{11} & A_{12}\\ A_{21} & A_{22}\end{pmatrix},
\]
where for the root system~$\mathsf{A}_2$ we have that
\begin{align*}
A_{11} &= 2 D_x^3 +2 w^1 D_x + w^1_x,\\
A_{12} &= -D_x^4 -w^1 D_x^2 + (3 w^2-2 w^1_x)\cdot D_x +(2 w^2_x-w^1_{xx}),\\
A_{21} &= D_x^4 + w^1 D_x^2 +3 w^2 D_x + w^2_x\\
A_{22} &= -\tfrac{2}{3} D_x^5 -\tfrac{4}{3} w^1 D_x^3 -2 w^1_x D_x^2 
 + (2 w^2_x-2 w^1_{xx}-\tfrac{2}{3} (w^1)^2)\cdot D_x 
 +\tfrac{1}{3}(3 w^2_{xx}-2 w^1_{xxx}-2 w^1 w^1_x).
\end{align*}
The bracket~\eqref{EqDogma} for $\smash{\hat{A}_k}$ equals
\begin{subequations}\label{A2SokBr}
\begin{align}
\ib{\vec{p}}{\vec{q}}{\hat{A}_k} &=   
  \boxed{p^1_x q^1-p^1 q^1_x} 
 +\bigl[p^1_{xx} q^2-p^2 q^1_{xx}\bigr]
 +\underline{\tfrac{2}{3} (p^2 q^2_{xxx}-p^2_{xxx} q^2)+\tfrac{2}{3} w^1 (p^2 q^2_x-p^2_x q^2)},\\
\ib{\vec{p}}{\vec{q}}{\hat{A}_k} &=
 \bigl[p^2_x q^1-p^1 q^2_x+2 (p^1_x q^2- p^2 q^1_x)\bigr]
  +\underline{p^2 q^2_{xx}-p^2_{xx} q^2}.
\end{align}
\end{subequations}
Consequently, not only the image of the entire operator~\eqref{SquareA2} 
is closed under the commutation,
but the image of the first column, of first order, is
itself closed under commutation (see, e.g., \cite{TMPhGallipoli} or the encyclopaedia~\cite{Ibragimov}, where only the first column of~\eqref{SquareA2} is presented).
However, the image of the second column of
$\square=\bigl(\square_1,\square_2\bigr)$ is not closed under the commutation.
Indeed, we box the individual bracket $\ib{\,}{\,}{\square_1}$ for the
$(2\times1)$-\/matrix operator~$\square_1$, and we underline the couplings
of components in the domain of~$\square_2$; under commutation, they hit 
both images of the first and second columns.

Performing the shift $w^2\mapsto w^2+\lambda$ of the second integral, and taking the velocity of the operator~$\smash{\hat{A}_k}$,
\[
\hat{A}_1^{(2)}\mathrel{{:}{=}}\frac{d}{d\lambda}{\Bigr|}_{\lambda=0}
   \bigl(\hat{A}_k\bigr),
\]
we obtain the `junior' \emph{Hamiltonian} operator
$\hat{A}_1^{(2)}=\left(\begin{smallmatrix}0 & 3D_x\\ 3D_x & 0\end{smallmatrix}
\right)$ that is compatible with the former.
Obviously, the bracket $\ib{\,}{\,}{\hat{A}_1^{(2)}}$ on the domain 
of~$\hat{A}_1^{(2)}$ vanishes identically.
(We note that the analogous operator 
$\hat{A}_1^{(1)}=\frac{d}{d\mu}{\bigr|}_{\mu=0}\bigl(\hat{A}_k\bigr)$, where
$w^1\mapsto w^1+\mu$, is not Hamiltonian at all.)

The pair $(\hat{A}_1^{(2)},\hat{A}_k)$ is the well\/-\/known bi\/-\/Hamiltonian structure for the Boussinesq equation
\begin{subequations}\label{EBous}
\begin{align}
w^1_t&=2w^2_x-w^1_{xx},\\
w^2_t&=-\tfrac{2}{3}w^1_{xxx}-\tfrac{2}{3}w^1w^1_x+w^2_{xx}.
\end{align}
\end{subequations}
Indeed, we have that
\[
\vec{w}_t=\hat{A}_1^{(2)}\,\frac{\delta}{\delta\vec{w}}
   \int\tfrac{1}{3}\Bigl[w^1w^2_x+\tfrac{1}{3}{(w^1_x)}^2
   -\tfrac{1}{6}{(w^1)}^2+{(w^2)}^2\Bigr]\,\Id x =
\hat{A}_k\,\frac{\delta}{\delta\vec{w}}
   \int w^2\,\Id x.
\]
Both densities, $w^1$ and $w^2$, are conserved on system~\eqref{EBous}.
The symmetry $\vec{w}_x=\hat{A}_k\,\frac{\delta}{\delta\vec{w}}\int w^1\,\Id x$ starts the second sequence of Hamiltonian flows in 
the Boussinesq hierarchy~$\mathfrak{A}$, see~\cite{TMPhGallipoli} and references therein.

The modified Boussinesq hierarchy~$\mathfrak{B}$ shares the two sequences of Hamiltonians with the Boussinesq hierarchy itself, by virtue of the Miura substitution $w=w[\gm]$ with $\gm=\gm[u]$. 
Namely, for any Hamiltonian~$\cH[w]$, the flows
\[
u_\tau=\frac{\delta\cH[\gm]}{\delta\gm},\qquad
\gm_\tau=-\frac{\delta\cH\bigl[\gm[u]\bigr]}{\delta u}
\]
belong to the modified hierarchy~$\mathfrak{B}$. The correlation between the two hierarchies, $\mathfrak{A}$ and~$\mathfrak{B}$, and the Hamiltonian structures,
\[
\hat{A}_k,\quad \hat{A}_1^{(2)}={(A_1)}^{-1},\quad A_k,\text{ and }
\hat{B}_1=\bigl(\ell_\gm^{(u)}\bigr)^*=B_1^{-1},\quad \hat{B}_{k'},\quad 
B_{k'},
\]
for their potential and nonpotential components are standard, see the diagram in~\cite[section~5.1]{Twente}.
The velocities~$u_\tau$ constitute the commutative subalgebra of Noether's symmetries of the 2D~Toda chain~\eqref{A2Toda}.

\section{The root system~$\mathsf{B}_2$}\noindent%
The Toda system is specified by the Cartan matrix
$K=\left(\begin{smallmatrix}\phantom{+}2& -2\\ -1 & \phantom{+}2\end{smallmatrix}\right)$:
\[
u_{xy}=\exp(2u-2v),\qquad v_{xy}=\exp(-u+2v).
\]
The integrals for it are of orders $1$ and $3$ with respect to the momenta:
\begin{subequations}\label{B2Int}
\begin{align}
w^1&=u_{xx} + 2 v_{xx} - 2 (v_x)^2 + 2 v_x u_x - u_x^2,\\
w^2&=v_{4x} + v_x (u_{xxx} - 2 v_{xxx}) + u_{xx} v_x (v_x - 2 u_x) \\ 
{}&\qquad  + v_{xx} (4 v_x u_x - 2 (v_x)^2 - (u_x)^2) + v_{xx} (u_{xx} - v_{xx}) \notag\\
{}&\qquad  + (v_x)^4 + (v_x)^2 (u_x)^2 - 2 (v_x)^3 u_x.\notag
\end{align}
\end{subequations}
Hence the Frobenius operator~\eqref{ISquare} is
\[
\square=\bigl(\square^1,\square^2\bigr)=
  \begin{pmatrix}\square_{11} & \square_{12}\\ \square_{21} & \square_{22}
  \end{pmatrix},\qquad\text{where 
   $\square^1=\binom{u_x+2D_x}{v_x+\tfrac{3}{2}D_x}$,}
\]
and  
\begin{align*}
\square_{12}&= 
 D_x^3 +\bigl(u_{xx}-u_x^2\bigr)\cdot D_x
  +\bigl(2 v_x^2 u_x-4 v_{xx} v_x+2 u_{xx} v_x-2 v_x u_x^2 +2 v_{xxx} \bigr)
;\\
\square_{22}&= 
 D_x^3 +v_x D_x^2
  +\bigl( 2 v_x u_x -v_x^2 -u_x^2 +v_{xx}+u_{xx} \bigr)\cdot D_x \\
{}&\qquad +\bigl( 4 v_x^2 u_x -2 v_{xx} v_x +2 u_{xx} v_x -2 v_x u_x^2
         +2 v_{xxx} -2 v_x^3 \bigr).
\end{align*}
The Hamiltonian operator 
$\smash{\hat{A}_k}=\begin{pmatrix} A_{11} & A_{12}\\ A_{21} & A_{22}\end{pmatrix}$
has the components
\begin{align*}
A_{11} &= 10 D_x^3 + 4 w^1 D_x + 2 w^1_x ,\\
A_{12} &= 3 D_x^5 +3 w^1 D_x^3 +6 w^1_x D_x^2 +(3 w^1_{xx} + 8 w^2)\cdot D_x 
   +6 w^2_x,\\
A_{21} &= 3 D_x^5 +3 w^1 D_x^3 +3 w^1_x D_x^2 +8 w^2 D_x +2 w^2_x,\\
A_{22} &= D_x^7 +2 w^1 D_x^5 +5 w^1_x D_x^4
   +(6 w^1_{xx}+6 w^2+ (w^1)^2)\cdot D_x^3 \\
{}&\qquad   +(4 w^1_{xxx}+3 w^1 w^1_x + 9 w^2_x)\cdot D_x^2 \\
{}&\qquad   +\bigl( w^1_{4x}+7 w^2_{xx} +  (w^1_x)^2 +4 w^1 w^2 + w^1 w^1_{xx}
    \bigr)\cdot D_x \\
{}&\qquad   +2\cdot( w^1_x w^2 + w^2_{xxx} + w^2_x w^1).
\end{align*}
Therefore the components of the brackets~\eqref{BothBrackets} 
for both~$\smash{\hat{A}_k}$ and~$\square$ are
\begin{align*}
\ib{\vec{p}}{\vec{q}}{\square}^1&= 
 2 (p^1 q^1_x - p^1_x q^1) 
+3 (p^1_{xx} q^2_x - p^2_x q^1_{xx})
+ (p^2_{4x} q^2_x - p^2_x q^2_{4x}) \\
{}&\qquad +w^1 (p^2_{xx} q^2_x  - p^2_x q^2_{xx})
+2w^2 (p^2 q^2_x - p^2_x q^2)
;\\
\ib{\vec{p}}{\vec{q}}{\square}^2&= 
 6 (p^2 q^1_x - p^1_x q^2)
+2 (p^1 q^2_x - p^2_x q^1)
+2 (p^2 q^2_{xxx} - p^2_{xxx} q^2) \\
{}&\qquad + (p^2_{xx} q^2_x - p^2_x q^2_{xx}) 
+2w^1 (p^2 q^2_x - p^2_x q^2).
\end{align*}
Similar to the case of~\eqref{A2SokBr}, the commutation of symmetries
that belong to the image of the second column~$\square^2$ of the
operator~$\square=\bigl(\square^1$,\ $\square^2\bigr)$ for~$\mathsf{B}_2$
hits the image of the first column~$\square^1$.

The `junior' Hamiltonian operator 
$\hat{A}_1^{(2)}=\frac{d}{d\lambda}{\bigr|}_{\lambda=0}\hat{A}_k$ 
is again obtained by taking the shift $w^2\mapsto w^2+\lambda$
in~$\hat{A}_k$:
\[
\hat{A}_1^{(2)} = \begin{pmatrix} 0 & 8D_x\\ 
8D_x & 6D_x^3+4w^1D_x+2w^1_x \end{pmatrix}.
\]
The new Hamiltonian operator is compatible with the old one.
The bracket on the domain of $\hat{A}_1^{(2)}$ is given by
\[
\ib{\vec{p}}{\vec{q}}{\hat{A}_1^{(2)}}^1=2(p^2_xq^2-p^2q^2_x),\qquad
\ib{\vec{p}}{\vec{q}}{\hat{A}_1^{(2)}}^2=0.
\]
Likewise to the root system~$\mathsf{A}_2$, the shift $w^1\mapsto w^1+\mu$
produces the operator
$\hat{A}_1^{(1)}=\frac{d}{d\mu}{\bigr|}_{\mu=0}\hat{A}_k$ 
which is not Hamiltonian.

The pair $\bigl(\hat{A}_1^{(2)},\hat{A}_k\bigr)$ determines the hierarchy of the KdV\/-\/type system
\begin{align*}
w^1_t&=6w^2_x,\\
w^2_t&=D_x\bigl(2w^2_{xx}+2w^1w^2\bigr).
\end{align*}
Here we have again that $\vec{w}_t=\hat{A}_k\frac{\delta}{\delta\vec{w}}
\int w^2\,\Id x$, and the translation $\vec{w}_x=\hat{A}_k\frac{\delta}{\delta\vec{w}}\int w^1\,\Id x$ starts the auxiliary sequence of flows.
The construction of the modified hierarchy is analogous to the previous case 
of the root system~$\mathsf{A}_2$, see~\cite{Twente} for details.

\section{The root system~$\mathsf{G}_2$}\noindent%
The Toda system for 
$K=\left(\begin{smallmatrix}\phantom{+}2& -1\\ -3 & \phantom{+}2\end{smallmatrix}\right)$ is
\begin{equation}\label{G2Toda}
u_{xy}=\exp(2u-v),\qquad v_{xy}=\exp(-3u+2v).
\end{equation}
The differential orders of the integrals w.r.t.\ the momenta equal $1$ and~$5$,
respectively:\label{G2Int}
\begin{align*}
w^1&=u_{xx}+\tfrac{1}{3} v_{xx}-u_x^2+u_x v_x-\tfrac{1}{3}v_x^2,\\
w^2&=u_{6x} -2 u_x u_{5x} +u_x v_{5x}
+10 u_{4x} u_x v_x-8 u_{4x} u_x^2-\tfrac{7}{3} u_{4x} v_x^2
+\tfrac{7}{3} u_{4x} v_{xx} \\
{}&\quad -\tfrac{5}{3} v_{4x} u_x v_x 
+\tfrac{2}{3} v_{4x} u_x^2
-\tfrac{1}{9} v_{4x} v_x^2
+\tfrac{1}{9} v_{4x} v_{xx}+\tfrac{10}{3} v_{4x} u_{xx}
+\tfrac{46}{3} u_{xxx} v_{xx} u_x
+20 u_{xxx} u_{xx} v_x \\
{}&\quad -40 u_{xxx} u_{xx} u_x
+\tfrac{10}{3} u_{xxx} v_{xxx}
-\tfrac{19}{3} u_{xxx} v_{xx} v_x
+16 u_{xxx} u_x^3
-18 u_{xxx} u_x^2 v_x
-\tfrac{1}{3} u_{xxx} v_x^3 \\
{}&\quad +\tfrac{14}{3} u_{xxx} u_x v_x^2
-\tfrac{19}{3} v_{xxx} u_{xx} v_x
+\tfrac{40}{3} v_{xxx} u_{xx} u_x 
-8 v_{xxx} u_x^3
-\tfrac{4}{9} v_{xxx} v_{xx} v_x
-\tfrac{13}{3} v_{xxx} v_{xx} u_x \\
{}&\quad +\tfrac{26}{3} v_{xxx} u_x^2 v_x
-2 v_{xxx} u_x v_x^2
+\tfrac{2}{9} v_x^3 v_{xxx}
+\tfrac{1}{18} v_{xxx}^2
-\tfrac{17}{3} u_{xx} v_{xx}^2 
+40 u_{xx}^2 u_x^2 
-28 u_{xx}^2 u_x v_x \\
{}&\quad -2 u_{xx} v_{xx} v_x^2+\tfrac{25}{6} u_{xx}^2 v_x^2
-16 u_{xx}^3
+\tfrac{1}{3} u_{xx} v_x^4-5 u_{xx} u_x v_x^3
+15 u_{xx} u_x^2 v_x^2
-12 u_{xx} u_x^3 v_x \\
{}&\quad +\tfrac{58}{3} u_{xx} v_{xx} u_x v_x
-34 u_{xx} v_{xx} u_x^2
+\tfrac{49}{3} u_{xx}^2 v_{xx}
+\tfrac{10}{3} v_{xx} v_x^3 u_x
-\tfrac{8}{27} v_{xx}^3
+12 v_{xx} v_x u_x^3 \\
{}&\quad -\tfrac{34}{3} v_{xx} v_x^2 u_x^2-\tfrac{2}{9} v_{xx} v_x^4
+\tfrac{25}{3} v_{xx}^2 u_x^2
-4 v_{xx}^2 u_x v_x+\tfrac{2}{3} v_{xx}^2 v_x^2
-3 v_{xx} u_x^4
-\tfrac{2}{3} u_x v_x^5
+\tfrac{2}{27} v_x^6 \\
{}&\quad +\tfrac{3}{2} u_x^4 v_x^2
+\tfrac{13}{6} u_x^2 v_x^4
-3 u_x^3 v_x^3.
\end{align*}
The Frobenius operator~$\square$ is 
\begin{equation}\label{G2Square}
\square=\bigl(\square^1,\square^2\bigr)=
  \begin{pmatrix}\square_{11} & \square_{12}\\ \square_{21} & \square_{22}
  \end{pmatrix},\qquad\text{where $\square^1=\binom{u_x+3D_x}{v_x+5D_x}$}
\end{equation}
and 
\begin{align*}
\square_{12}&= 
2 D_x^5
+u_x D_x^4
+\Bigl(15 u_x v_x-14 u_x^2-5v_x^2+14u_{xx}+5v_{xx}\Bigr)\cdot D_x^3 \\
{}&+\Bigl(8v_{xxx}-16 v_{xx} v_x+10u_x^2 v_x+\tfrac{82}{3} v_{xx} u_x
-44u_{xx} u_x
-\tfrac{10}{3} u_x v_x^2+24u_{xx} v_x-8u_x^3+26 u_{xxx}\Bigr)\cdot D_x^2 \\
{}&+\Bigl(12 u_x^3 v_x-16 v_{xxx} v_x-38 u_{xxx} u_x-4 u_x^2 v_x^2
+4 u_{xx} v_x^2+24 u_{xxx} v_x+44 u_{xx} v_{xx} \\
{}&\qquad +2 v_{xx} u_x^2
+\tfrac{70}{3} v_{xxx} u_x+10 u_{xx} u_x^2+21 u_{4x}+8 v_{4x}-16 v_{xx}^2
-14 u_{xx} u_x v_x \\
{}&\qquad +\tfrac{4}{3} v_{xx} u_x v_x-51 u_{xx}^2-9 u_x^4\Bigr)\cdot D_x \\
{}&+\Bigl(15 u_x^3 v_x^2-23 u_{xx} u_x v_x^2-\tfrac{44}{3} v_{xxx} u_x v_x
+\tfrac{4}{3} u_x v_x^4+18 u_{xx} u_x^2 v_x
-12 v_{xx} u_x^2 v_x \\
{}&\qquad +\tfrac{46}{3} v_{xx} u_x v_x^2+28 u_{xx} v_{xx} u_x
-4 u_{xx} v_{xx} v_x+10 u_{xxx} u_x v_x+12 u_{5x}+3 v_{5x} \\
{}&\qquad +u_{xxx} v_x^2
-20 u_{4x} u_x+4 v_{xx}^2 v_x+\tfrac{40}{3} v_{4x} u_x-6 v_{4x} v_x
+9 u_{4x} v_x-72 u_{xxx} u_{xx} \\
{}&\qquad -20 v_{xxx} v_{xx}+26 v_{xxx} u_{xx}
-8 u_x^2 v_x^3+6 u_{xx} v_x^3-8 u_{xx}^2 u_x-4 v_{xx} v_x^3 \\
{}&\qquad +26 u_{xxx} v_{xx}+2 v_{xxx} v_x^2-8 u_{xxx} u_x^2-\tfrac{40}{3} v_{xx}^2 u_x
-9 u_x^4 v_x-3 u_{xx}^2 v_x+14 v_{xxx} u_x^2\Bigr),
\end{align*}
\begin{align*}
\square_{22}&= 
3 D_x^5
+\Bigl(20 u_x v_x-20 u_x^2+\tfrac{23}{3} v_{xx}
  -\tfrac{23}{3} v_x^2+20 u_{xx}\Bigr)\cdot D_x^3 \\
{}&+\Bigl(40 u_{xxx}-2 u_x v_x^2-80 u_{xx} u_x-\tfrac{73}{3} v_{xx} v_x
 +2 u_x^2 v_x+40 v_{xx} u_x+\tfrac{37}{3} v_{xxx}+38 u_{xx} v_x
 -\tfrac{1}{3} v_x^3\Bigr)\cdot D_x^2 \\
{}&+\Bigl(38 u_{xxx} v_x-25 v_{xx}^2-\tfrac{73}{3}v_{xxx} v_x-68 u_{xxx}
 u_x-23u_x^2 v_x^2+36 u_x^3 v_x+9u_{xx} v_x^2 \\
{}&\qquad +66u_{xx} v_{xx}+2 v_{xx} u_x^2
 +34 v_{xxx} u_x+36 u_{xx} u_x^2-86 u_{xx}^2-18 u_x^4
 +34 u_{4x} \\
{}&\qquad +\tfrac{37}{3} v_{4x}-44 u_{xx} u_x v_x+2 v_{xx} u_x v_x
 -\tfrac{1}{3} v_x^4+5 u_x v_x^3\Bigr)\cdot D_x \\
{}&+\Bigl(36 u_{xx} u_x^2 v_x-22 v_{xxx} u_x v_x-\tfrac{2}{3} v_x^5
+36 u_x^3 v_x^2+7 u_x v_x^4-46 u_{xx} u_x v_x^2-18 v_{xx} u_x^2 v_x \\
{}&\qquad +23 v_{xx} u_x v_x^2+40 u_{xx} v_{xx} u_x-6 u_{xx} v_{xx} v_x+4 u_{xxx} u_x v_x+20 u_{5x}
+\tfrac{14}{3} v_{5x} \\
{}&\qquad +3 u_{xxx} v_x^2-40 u_{4x} u_x+6 v_{xx}^2 v_x
+20 v_{4x} u_x-9 v_{4x} v_x+18 u_{4x} v_x-120 u_{xxx} u_{xx} \\
{}&\qquad -\tfrac{95}{3} v_{xxx} v_{xx}+40 v_{xxx} u_{xx}-25 u_x^2 v_x^3
+12 u_{xx} v_x^3-6 v_{xx} v_x^3+40 u_{xxx} v_{xx}+3 v_{xxx} v_x^2 \\
{}&\qquad -20 v_{xx}^2 u_x-18 u_x^4 v_x-14 u_{xx}^2 v_x+20 v_{xxx} u_x^2\Bigr).
\end{align*}
Using the fact that the coefficients $A_{ij}$ of the 
Hamiltonian operator 
$\smash{\hat{A}_k}=\begin{pmatrix} A_{11} & A_{12}\\ A_{21} & A_{22}\end{pmatrix}$
are differential functions of the integrals~$w$, 
we deduce for~$\mathsf{G}_2$ that\footnote{%
Here part (\ref{DependOnW}) of our main theorem reveals its true power:
a verification for~$\mathsf{G}_2$ that the bracket $\ib{\,}{\,}{\square}$, which depends on the fields~$u$ through the integrals~$w^i$, satisfies 
equality~\eqref{OplusBG2}, results in a 50~Mb size expression.}
\begin{align*}
A_{11} &= \tfrac{14}{3} D_x^3 +2 w^1 D_x+w^1_x,\\
A_{12}&=3 D_{x}^7 +\tfrac{68}{3} w^1 D_{x}^5 +\tfrac{245}{3} w^1_x D_{x}^4 
 +\bigl((w^1)^2 +\tfrac{395}{3} w^1_{xx}\bigr)\cdot D_{x}^3 
 +\bigl(\tfrac{368}{3} w^1_{xxx} -28 w^1 w^1_x\bigr)\cdot D_{x}^2 \\
{}&\qquad +\bigl(6 w^2 +\tfrac{191}{3} w^1_{4x} -62 w^1_{xx}  w^1-62 (w^1_x)^2\bigr)\cdot D_x 
 +\bigl(5 w^2_x +\tfrac{41}{3} w^1_{5x} -32 w^1 w^1_{xxx} -96 w^1_{xx} w^1_x\bigr),\\
A_{21}&=3 D_x^7 +\tfrac{68}{3} w^1 D_x^5 
  +\tfrac{95}{3} w^1_x D_x^4  
  +\bigl((w^1)^2+\tfrac{95}{3} w^1_{xx} \bigr)\cdot D_x^3
  +\bigl(9 w^1_{xxx} +34 w^1 w^1_x\bigr)\cdot D_x^2
  +6 w^2 D_x + w^2_x,\\
A_{22}&=2 D^{11}_{x} +30 w^1 D^9_{x} +135 w^1_x D^8_{x} 
+\bigl(414 w^1_{xx} +\tfrac{338}{3} (w^1)^2\bigr)\cdot D^7_{x} 
+\bigl(819 w^1_{xxx} +\tfrac{2366}{3} w^1 w^1_x\bigr)\cdot D^6_{x} \\
{}&\quad +\bigl(1119 w^1_{4x} +4 w^2+\tfrac{4870}{3} w^1_{xx} w^1
  +\tfrac{1846}{3} (w^1_x)^2+8 (w^1)^3\bigr)\cdot D^5_{x} \\
{}&\quad +\bigl(1065 w^1_{5x} +10 w^2_x+\tfrac{6260}{3} w^1 w^1_{xxx} 
 +1220 w^1_{xx} w^1_x+60 w^1_x (w^1)^2\bigr)\cdot D^4_{x} \\
{}&\quad +\Bigl(699 w^1_{6x} +26 w^2_{xx} +\tfrac{5402}{3} w^1_{4x}  w^1
 +36 w^1 w^2+\tfrac{1096}{3} w^1_{xxx}  w^1_x-\tfrac{652}{3} (w^1_{xx})^2
 -428 w^1_{xx}  (w^1)^2 \\
{}&\qquad -88 (w^1_x)^2 w^1+54 (w^1)^4\Bigr)\cdot D^3_{x} \\
{}&\quad +\Bigl(303 w^1_{7x} +29 w^2_{xxx} +\tfrac{3026}{3} w^1_{5x}  w^1+54 w^2_x w^1
 -\tfrac{518}{3} w^1_{4x}  w^1_x+54 w^1_x w^2-\tfrac{5722}{3} w^1_{xxx} w^1_{xx} \\
{}&\qquad -702 w^1_{xxx}  (w^1)^2-1908 w^1_{xx}  w^1_x w^1-252 (w^1_x)^3
 +324 w^1_x (w^1)^3\Bigr)\cdot D^2_{x} \\
{}&\quad +\Bigl(78 w^1_{8x} +15 w^2_{4x} +328 w^1_{6x}  w^1+58 w^2_{xx}  w^1
 -\tfrac{560}{3} w^1_{5x}  w^1_x+36 w^2_x w^1_x-\tfrac{4030}{3} w^1_{4x} w^1_{xx} \\
{}&\qquad -514 w^1_{4x}  (w^1)^2+54 w^2 w^1_{xx} -36 w^2 (w^1)^2-1086 (w^1_{xxx})^2
 -2000 w^1_{xxx}  w^1_x w^1 \\
{}&\qquad -1652 (w^1_{xx})^2 w^1-1120 w^1_{xx} (w^1_x)^2
 +396 w^1_{xx}  (w^1)^3+540 (w^1_x)^2 (w^1)^2\Bigr)\cdot D_x 
\end{align*}
\begin{align*}\phantom{A_{22}}
{}&\quad +\Bigl(9 w^1_{9x} +3 w^2_{5x} +\tfrac{140}{3} w^1_{7x}  w^1+20 w^2_{xxx}  w^1
 -\tfrac{148}{3} w^1_{6x}  w^1_x+20 w^2_{xx}  w^1_x -\tfrac{994}{3} w^1_{5x}  w^1_{xx} \\
{}&\qquad -138 w^1_{5x} (w^1)^2+18 w^2_x w^1_{xx} 
 -18 w^2_x (w^1)^2-678 w^1_{4x}  w^1_{xxx} -708 w^1_{4x}  w^1 w^1_x+18 w^2 w^1_{xxx} \\
{}&\qquad -36 w^2 w^1_x w^1-w^1_{xxx}\cdot\bigl(1424 w^1_{xx}  w^1+524 (w^1_x)^2-144 (w^1)^3\bigr)
 -680 (w^1_{xx})^2 w^1_x \\
{}&\qquad +648 w^1_{xx}  w^1_x (w^1)^2+216 (w^1_x)^3 w^1\Bigr).
\end{align*}
Next, we calculate the bracket~\eqref{BothBrackets} 
on the domain of~\eqref{G2Square} using formula~\eqref{EqDogma}:
\begin{align*}
\ib{\vec{p}}{\vec{q}}{\square}^1&=
\bigl[p^1_x q^1-p^1 q^1_x\bigr]
+\bigl[\tfrac{41}{3} (p^1_{5x} q^2 -p^2 q^1_{5x} )
+\tfrac{14}{3} (p^1_{4x}  q^2_x - p^2_x q^1_{4x} )
+\tfrac{14}{3} (p^1_{xxx}  q^2_{xx} - p^2_{xx}  q^1_{xxx} ) \\
{}&\quad +9 (p^2_{xxx} q^1_{xx} - p^1_{xx} q^2_{xxx} )
+32 w^1 (p^2 q^1_{xxx} - p^1_{xxx} q^2 )
+34 w^1 (p^2_x q^1_{xx} - p^1_{xx} q^2_x )\bigr] \\
{}&\quad +9 (p^2_{9x}  q^2 - p^2 q^2_{9x} )
+3 (p^2_{8x}  q^2_x - p^2_x q^2_{8x} ) \\
{}&\quad +\tfrac{140}{3} w^1 (p^2_{7x} q^2 - p^2 q^2_{7x} ) 
+3 (p^2_{7x}  q^2_{xx} - p^2_{xx}  q^2_{7x} )
+6 (p^2_{xxx}  q^2_{6x} - p^2_{6x}  q^2_{xxx} ) \\
{}&\quad +376 w^1_x (p^2_{6x} q^2 - p^2 q^2_{6x} ) 
+\tfrac{4}{3} w^1 (p^2_x q^2_{6x} - p^2_{6x} q^2_x )
+\tfrac{62}{3} w^1 (p^2_{5x} q^2_{xx} - p^2_{xx} q^2_{5x} ) \\
{}&\quad +\tfrac{304}{3} w^1_x (p^2_{5x} q^2_x - p^2_x q^2_{5x} ) 
+\tfrac{2834}{3} w^1_{xx} (p^2_{5x} q^2 - p^2 q^2_{5x} )
+138 (w^1)^2 (p^2 q^2_{5x} - p^2_{5x} q^2 ) \\
{}&\quad +\tfrac{248}{3}  w^1_x (p^2_{4x} q^2_{xx} - p^2_{xx} q^2_{4x} ) 
+\tfrac{640}{3}  w^1_{xx} (p^2_{4x} q^2_x - p^2_x q^2_{4x} )
+\tfrac{4184}{3}  w^1_{xxx} (p^2_{4x} q^2 - p^2 q^2_{4x} ) \\
{}&\quad +44 w^1 (p^2_{xxx} q^2_{4x} - p^2_{4x} q^2_{xxx} )
+672 w^1 w^1_x (p^2 q^2_{4x}  - p^2_{4x} q^2)
+176 (w^1)^2 (p^2_x q^2_{4x}  - p^2_{4x} q^2_x ) \\
{}&\quad +\tfrac{830}{3} w^1_{xx} (p^2_{xxx} q^2_{xx} - p^2_{xx} q^2_{xxx} )
+\tfrac{1060}{3} w^1_{xxx} (p^2_{xxx}  q^2_x - p^2_x q^2_{xxx} )
+\tfrac{4022}{3} w^1_{4x} (p^2_{xxx}  q^2 - p^2 q^2_{xxx} ) \\
{}&\quad +452 (w^1_x)^2 (p^2 q^2_{xxx} - p^2_{xxx}  q^2 )
+144 (w^1)^3 (p^2_{xxx} q^2 - p^2 q^2_{xxx} )
+18 w^2 (p^2_{xxx}  q^2 - p^2 q^2_{xxx} )\\
{}&\quad +26 (w^1)^2 (p^2_{xx}  q^2_{xxx}  - p^2_{xxx}  q^2_{xx} ) 
+1352 w^1_{xx}  w^1 (p^2 q^2_{xxx} - p^2_{xxx} q^2 )\\
{}&\quad +576  w^1 w^1_x (p^2_x q^2_{xxx}  - p^2_{xxx} q^2_x )
+\tfrac{368}{3} w^1_{4x} (p^2_{xx} q^2_x - p^2_x q^2_{xx} ) 
+1360 w^1 w^1_{xxx} (p^2 q^2_{xx}  - p^2_{xx}  q^2 ) \\
{}&\quad +1592 w^1_{xx} w^1_x (p^2 q^2_{xx} - p^2_{xx} q^2 ) 
+648 w^1_x (w^1)^2 (p^2_{xx} q^2  - p^2 q^2_{xx} ) 
+68  (w^1_x)^2 (p^2_x q^2_{xx}  - p^2_{xx}  q^2_x ) \\
{}&\quad +772 w^1_{5x} (p^2_{xx} q^2  - p^2 q^2_{xx} ) 
+36 w^2_x (p^2_{xx} q^2  - p^2 q^2_{xx} ) 
+36 (w^1)^3 (p^2_{xx} q^2_x - p^2_x q^2_{xx} ) \\
{}&\quad +584 w^1_{xx} w^1 (p^2_x q^2_{xx} - p^2_{xx} q^2_x ) 
+\tfrac{772}{3} w^1_{6x} (p^2_x q^2 - p^2 q^2_x )
+1020 (w^1_{xx})^2 (p^2 q^2_x  - p^2_x q^2 ) \\
{}&\quad +38 w^2_{xx} (p^2_x q^2 - p^2 q^2_x ) 
+1360 w^1_{xxx}  w^1_x (p^2 q^2_x  - p^2_x q^2 )
+680 w^1_{4x}  w^1 (p^2 q^2_x  - p^2_x q^2 ) \\
{}&\quad +36 w^1 w^2 (p^2 q^2_x - p^2_x q^2 ) 
+648 w^1_{xx}  (w^1)^2 (p^2_x q^2  - p^2 q^2_x )
+648 (w^1_x)^2 w^1 (p^2_x q^2  - p^2 q^2_x ) ;\\
\ib{\vec{p}}{\vec{q}}{\square}^2&=
p^2_x q^1 - p^1 q^2_x
+5 (p^1_x q^2 - p^2 q^1_x ) 
+3 (p^2_{5x} q^2 - p^2 q^2_{5x} )
+ (p^2_{xx}  q^2_{xxx} -p^2_{xxx}  q^2_{xx} ) \\
{}&\quad +20 w^1 (p^2_{xxx} q^2 - p^2 q^2_{xxx} )
+40 w^1_x (p^2_{xx} q^2 - p^2 q^2_{xx} )
+2 w^1 (p^2_{xx} q^2_x - p^2_x q^2_{xx} ) \\
{}&\quad +38 w^1_{xx} (p^2_x q^2 - p^2 q^2_x )
+18 (w^1)^2 (p^2 q^2_x  - p^2_x q^2 ).
\end{align*}
The equalities
\[
\bigl[\hat{A}_k(\vec{p}),\hat{A}_k(\vec{q})\bigr]=
\hat{A}_k\Bigl(\cEv_{\hat{A}_k(\vec{p})}(\vec{q})-
 \cEv_{\hat{A}_k(\vec{q})}(\vec{p})+\ib{\vec{p}}{\vec{q}}{\hat{A}_k}\Bigr)
\]
and
\begin{equation}\label{OplusBG2}
\bigl[\square(\vec{p}),\square(\vec{q})\bigr]=
\square\Bigl(\cEv_{\square(\vec{p})}(\vec{q})-
 \cEv_{\square(\vec{q})}(\vec{p})+\ib{\vec{p}}{\vec{q}}{\square}\Bigr)
\end{equation}
hold, where $\ib{\vec{p}}{\vec{q}}{\square}=
 \ib{\vec{p}}{\vec{q}}{\hat{A}_k}$ for any $\vec{p}$,\ $\vec{q}(x,[w])$.
This yields all the commutation relations between symmetries 
$\vph=\square(\cdot)$ of the 2D~Toda chain~\eqref{G2Toda}
associated with the root system~$\mathsf{G}_2$.

Finally, we pass to the KdV\/-\/type hierarchy.
The deformation $\frac{d}{d\lambda}{\bigr|}_{\lambda=0}\hat{A}_k$
under $w^2\mapsto w^2+\lambda$ determines the `junior' Hamiltonian operator
\[
\hat{A}_1^{(2)}=\begin{pmatrix} 0 & 6D_x \\ 6D_x &
4D_x^5+36w^1D_x^3+54w^1_xD_x^2+\bigl(54w^1_{xx}-36{(w^1)}^2\bigr)\cdot D_x 
  +18w^1_{xxx}-36w^1w^1_x
\end{pmatrix}.
\]
It is compatible with the `senior' operator~$\hat{A}_k$;
the bracket on its domain is given through
\[
\ib{\vec{p}}{\vec{q}}{\hat{A}_1^{(2)}}^1=
  36w^1\bigl(p^2_xq^2-p^2q^2_x\bigr) +18\bigl(p^2q^2_{xxx}-p^2_{xxx}q^2\bigr),
\qquad
\ib{\vec{p}}{\vec{q}}{\hat{A}_1^{(2)}}^2=0.
\]
Applying the Hamiltonian operator $\hat{A}_k$ to $\frac{\delta}{\delta\vec{w}}
\int w^2\,\Id x=\binom{0}{1}$, we obtain the KdV\/-\/type system
\begin{align*}
w^1_t&=D_x\Bigl(\tfrac{41}{3}w^1_{4x}-32w^1w^1_{xx}-32{(w^1_x)}^2+5w^2\Bigr),\\
w^2_t&=D_x\Bigl(3w^2_{4x} +9w^1_{8x} +20w^1w^2_{xx} +18w^1_{xx}w^2
  -18{(w^1)}^2w^2 +\tfrac{140}{3}w^1_{6x}w^1 \\
{}&{}\qquad  -96w^1_{5x}w^1_x
  -\tfrac{706}{3}w^1_{4x}w^1_{xx} -\tfrac{664}{3}{(w^1_{xxx})}^2
  -138w^1_{4x}{(w^1)}^2 -432w^1_{xxx}w^1_xw^1 \\
{}&{}\qquad -496{(w^1_{xx})}^2w^1
  -92w^1_{xx}{(w^1_x)}^2  +144w^1_{xx}{(w^1)}^3 +108{(w^1_x)}^2{(w^1)}^2\Bigr).
\end{align*}
(The second equation in this system can be simplified by adding to~$w^2$ a scaling\/-\/homogeneous differential polynomial in~$w^1$ and thus cancelling some irrelevant terms.)
The Hamiltonian $\int w^1\,\Id x$ starts the auxiliary sequence of flows by the translation along~$x$.
The corresponding modified KdV\/-\/type hierarchy, and the Hamiltonian
structures $\hat{B}_1=B_1^{-1}$, $\hat{B}_{k'}$, and~$B_{k'}$ for it, are introduced in a standard way~\cite{Twente}.

\section*{Discussion}
The geometric method for derivation of completely integrable KdV\/-\/type hierarchies, which is illustrated in this note, is the most straightforward and efficient, to the best of our knowledge. Unlike in the fundamental paper~\cite{DSViniti84}, which is based on algebraic considerations, the KdV\/-\/type systems are derived here, from the very beginning, in the bi\/-\/Hamiltonian but not in the Lax form. Second, the formalism of pseudodifferential operators is not required. We pass from the root systems directly to the 2D~Toda chains and then to the Poisson structures, skipping over the matrix representations of the semi\/-\/simple Lie algebras~$\mathfrak{g}$ and 
the techniques for fixing the gauges. \textit{A posteriori}, 
the Lax pairs for the KdV\/-\/type systems, 
which are bi\/-\/Hamiltonian\footnote{%
Let us finally remark that $k$~may be greater than~$2$: for instance, we have that $k=3$ for the Kaup\/--\/Boussinesq system, see~\cite{Twente}, and two `junior' structures precede~it.}
w.r.t.\ the operators $\hat{A}_1^{(2)}$ and~$\hat{A}_k$, can be obtained in a standard way, $R_t+[\ell_F,R]=0$, 
by using the recursion $R=\hat{A}_k\circ{\bigl(\hat{A}_1^{(2)}\bigr)}^{-1}$ and the linearization~$\ell_F$ of the right\/-\/hand sides~$F$ 
of the KdV\/-\/type systems~$\vec{w}_t=F$.
We argue, however, that this may not be a drawback of our concept.
Indeed, for $\mathfrak{g}$ semi\/-\/simple, the ambient $r$\/-\/component 2D~Toda chain is \emph{exactly solvable}. Therefore, in principle, the modified KdV\/-\/type flows should be lifted first to the bundles with $2r$-\/dimensional fibres, whose sections $\boldsymbol{f}(x)=(f_1,\ldots,f_r)$,
$\boldsymbol{g}(y)=(g_1,\ldots,g_{\bar{r}})$ determine the general solutions of the 2D~Toda chains. We see that the liftings of the mKdV\/-\/type hierarchies determine the evolution of these Cauchy data. It is well known that the Krichever\/--\/Novikov equation appears in this context for the root system~$\mathsf{A}_1$. In our opinion, the posing of the problem of 
integrability by the inverse scattering is more appropriate for that class of evolution equations, which may be not bi\/-\/Hamiltonian.

The Lax approach of~\cite{DSViniti84} becomes truely inevitable if the algebras~$\mathfrak{g}$ are Kac\/--\/Moody algebras, 
and the Cartan matrices~$K$ become degenerate. 
In this case, our cut\/-\/through does not work without serious modifications. This will be discussed elsewhere.

There is one more thing that we lack, it seems, due to the conscious refuse of the algebraic language in favour of the geometry of~PDE. Namely, this is an explanation of the way the `junior' Hamiltonian operators are obtained through the deformations of the canonical operators~$\hat{A}_k$, see~\eqref{Quattro}. The preference of certain shift directions for the integrals must have some explanation in terms of the cohomology of the $W$-\/algebras for the KdV\/-\/type equations at hand. Also, we do not know a group\/-\/theoretic origin for the resolvability of the Magri schemes, that is, the existence of the next Hamiltonian at each step (equivalently, the vanishing of the first Poisson cohomology with respect to the `junior' operator~$\hat{A}_1^{(2)}$). The most prominent technique for proving that is the construction of Gardner's deformations for the Drin\-fel'd\/--\/So\-ko\-lov systems.

\subsection*{Acknowledgements}
The authors thank B.\,A.\,Dub\-ro\-vin for illuminating discussions,
V.\,V.\,So\-ko\-lov for drawing A.\,K.'s attention to the 
paper~\cite{Shabat95}, and E.\,V.\,Ferapontov for helpful advice.
The authors are grateful to the organizing committee of the 4th International workshop ``Group analysis of differential equations and integrable systems.''
A.\,K. thanks the organizing committee of the 37th International workshop
``Seminar Sophus Lie'' (Paderborn, 
2009) for financial support and warm hospitality.
This work has been partially supported by the European Union through
the FP6 Marie Curie RTN \emph{ENIGMA} (Contract
no.\,MRTN-CT-2004-5652), the European Science Foundation Program
{MISGAM}, and by NWO VENI grant 639.031.623.
A part of this research was done while A.\,K.\ was visiting
at~$\smash{\text{IH\'ES}}$ and~SISSA, whose financial support is gratefully acknowledged.

\end{document}